\documentclass[10pt,letterpaper,twoside,twocolumn,english,showpacs]{revtex4}
\usepackage{times}
\usepackage[OT1]{fontenc}
\usepackage[latin1]{inputenc}
\usepackage{graphicx}
\usepackage{amssymb}

\makeatletter

\providecommand{\tabularnewline}{\\}

\usepackage{babel}
\makeatother
\begin{document}

\title{Wannier states and Bose-Hubbard parameters for 2D optical lattices}

\author{P. Blair Blakie}
\email{bblakie@physics.otago.ac.nz}
\affiliation{Department of Physics, University of Otago, P.O. Box 56, Dunedin, New Zealand}
\affiliation{Electron and Optical Physics Division, National Institute of Standards
and Technology, Gaithersburg, MD 20899-8410.}
\author{Charles W. Clark}
\affiliation{Electron and Optical Physics Division, National Institute of Standards
and Technology, Gaithersburg, MD 20899-8410.}

\date{\today}

\begin{abstract}
We consider the physical implementation of a 2D optical lattice with
schemes involving 3 and 4 light fields. We illustrate the wide range
of geometries available to the 3 beam lattice, and compare the general
potential properties of the two lattice schemes. Numerically calculating
the band structure we obtain the Wannier states and evaluate the parameters
of the Bose-Hubbard models relevant to these lattices. Using these
results we demonstrate lattices that realize Bose-Hubbard models with
2, 4, or 6 nearest neighbors, and quantify the extent that these different
lattices effect the superfluid to Mott-insulator transition.
\end{abstract}

\pacs{03.75.Hh, 32.80.Lg, 03.75.Lm}

\maketitle

\section{Introduction}

There has been considerable recent interest in the study of ultra-cold
bosonic atoms in optical lattices. The many favorable attributes of
optical lattices, such as the absence of defects, low noise level,
and high degree of experimental control, are ideal for precise quantum
manipulation. As such, this system features in theoretical proposals
for quantum computing (e.g. see \cite{Deutsch1998a,Calarco2000a}),
simulating many-body systems \cite{Jaksch2003a}, and studying quantum-phase
transitions \cite{Jaksch1998a}. Beyond bosons, work by Hofstetter
\emph{et al.} \cite{Hofstetter2002a} suggests \emph{}using an optical
lattice to facilitate the superfluid transition in a degenerate gas
of fermionic atoms \cite{Hofstetter2002a}.

Experimental studies with optical lattices have demonstrated many
impressive results, including number-squeezing an initially coherent
Bose-field using a 1D lattice \cite{Orzel2001a}; the superfluid to
Mott-insulator transition in a 3D lattice \cite{Greiner2002a}; the
collapse and revival of coherence in a matter-wave field \cite{Greiner2002b};
controlled collisions using state selective lattices \cite{Mandel2003a}.
All of these experiments are carried out at sufficiently low temperatures
and in deep enough lattices that the constituent bosons occupy only
the lowest band, and are well described by a tight-binding Bose-Hubbard
model \cite{Fisher1989a}. In this model the Bose-field is decomposed
into a set of localized Wannier states, one at each site of the lattice,
and the Hamiltonian is characterized by a few parameters: (i) the
number of nearest neighbors surrounding each lattice site, (ii) the
tunneling strength between nearest neighbors, and (iii) the strength
of the interaction between particles at the same site. 

The ground state of the Bose-Hubbard Hamiltonian exhibits a quantum
phase transition between superfluid and Mott-insulating states, that
depends on the relative size of the aforementioned parameters. The
transition to an insulating state has been suggested as a method for
preparing a fiducial state of precisely one atom per site, suitable
for quantum information processing. Another important application
of optical lattices is to produce tightly confining potentials suitable
for realizing low dimensional gases. For instance, in \cite{Burger2002a}
a 1D lattice was used to investigate Bose-condensation in a quasi-2D
gas, and experiments with 2D lattices \cite{NIST2003a,Moritz2003a}
have produced correlated 1D gases in the elongated potential tubes
at each lattice site. For these systems the tunneling parameters and
Wannier states characterize the time scale over which each potential
is isolated from neighboring sites, and transverse wavefunction for
the particles respectively.

To date most theory, and experiments for quantum-degenerate Bose-gases
have used cubic lattices, where the optical potential is produced
by orthogonal sets of counter propagating light fields. For this case
the lattice potential is separable, and the non-interacting band structure
is given by the Mathieu spectrum \cite{HandBkMathFns}. In this paper
we explore the structure and properties of a more general class of
two-dimensional optical lattices for regimes relevant to current experiments.
We compare and contrast the usual square separable lattice (arising
from two sets of independent laser fields) with a more general non-separable
lattice produced from three interfering light fields. To characterize
the appropriate Bose-Hubbard models for the lattices considered we
numerically evaluate the localized Wannier states from band structure
calculations. We discuss favorable properties of the 3 beam lattice
that make it more ideal for investigating the strongly correlated
regime. In particular, we show that certain 3 beam lattice configurations
provide tighter on-site confinement, control over the number of nearest
neighbors, and significantly reduced tunneling between sites compared
to the counter propagating 4 beam arrangement. These properties of
the 3 beam lattice lead to the superfluid to Mott-insulator transition
occurring for much shallower lattices, and would facilitate taking
experiments deeper into the Mott-insulating regime. We also show that
by using non-counter propagating light fields in the 4 beam lattice,
the depth at which the Mott-transition occurs can be significantly
reduced, albeit with the requirement of an additional light field. 

Conservative two dimensional optical lattices are an important tool
for the emerging field of quantum atom-optics, with numerous practical
applications \cite{Greiner2001a,Moritz2003a,NIST2003a,Jaksch2003a}.
We see the calculations presented in this paper as serving two important
purposes. First, for people studying the idealized Bose-Hubbard model,
these results will elucidate the typical range of Bose-Hubbard parameters
accessible, and demonstrate the flexibility of the optical lattice
system. Second, we believe these results will be important in design
considerations for future experiments with ultra-cold gases in optical
lattices.

\section{Formalism}

\subsection{Optical Lattice\label{sub:Optical-Lattice}}

\begin{figure}[!htb]
\includegraphics[%
  width=3.5in,
  keepaspectratio]{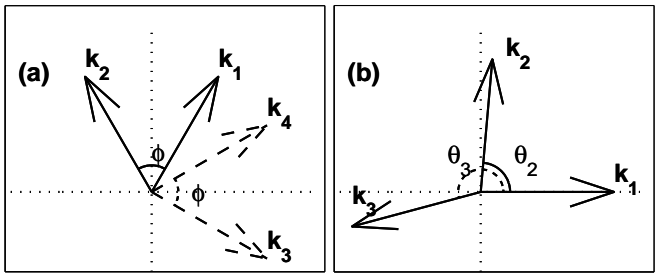}

\caption{\label{cap:LASERdiag} Wave vectors of the laser configurations used
to generate a 2D optical lattice. (a) 4 beam lattice formed by two
independent sets of plane-wave light fields. Different line types
are used to differentiate the two pairs. (b) 3 beam lattice produced
by three interfering plane-wave fields. The angles are discussed in
the text.}
\end{figure}

We consider an optical lattice produced from a set of monochromatic
plane-waves with propagation directions characterized by their wave
vectors $\{\mathbf{k}_{j}\}$. 
To simplify our discussion of 2D optical lattices we take these propagation directions to all lie in a single plane, which we designate
as the $x$-$y$ plane. For simplicity we consider all light fields
to be linearly polarized along $z$ and of equal intensity, with the
focus of this work on the control of lattice geometry obtained by
varying the light field propagation directions. The lattice potential
experienced by the atoms is proportional to the real part of the dynamic
polarizability $\alpha$ %
\footnote{The imaginary part of the dynamic polarizability characterizes the
dissipative part of the atom-light interaction (i.e. spontaneous emission),
and we assume that the detuning from atomic resonance is sufficiently
large that this can be neglected on the duration of experiments (e.g.
see \cite{Grimm2000a}).%
}, and the light field intensity $I_{0}$ %
\footnote{For a useful review of optical lattices and optical dipole forces
we refer to reader to the review articles \cite{Guidoni1999a,Grimm2000a}.%
}. Below we discuss the two different laser configurations we compare
in this paper.

\paragraph*{4 Beam Square Lattice}

In Fig. \ref{cap:LASERdiag}(a) we show the geometry used to realize
the 2D optical lattice we refer to as the 4 beam (4B) square lattice.
This type of lattice (and its 3D generalization) has been extensively
employed in recent experiments for the case where each pair of lasers
is counter propagating (i.e. with $\phi=180^{\circ}$) \cite{Greiner2001a,Greiner2002a}.
The two pairs of light fields are made independent from each other
by detuning the common frequency in one pair of fields from that of
the other pair. Typically, a negligible frequency difference compared
to the optical frequency is required to achieve independence (e.g.
see \cite{Peil2003a}), so to a good approximation all wave-vectors
are the same magnitude $|\mathbf{k}_{j}|\approx k\equiv2\pi/\lambda,$
where $\lambda$ is the optical wavelength. The resulting potential
is a superposition of two perpendicular 1D lattices (one arising from
each pair), and in the counter propagating case the lattice site separation
is $\lambda/2$. As illustrated in Fig. \ref{cap:LASERdiag}(a), we
have generalized this arrangement from counter propagating beams to
consider the pairs of light fields propagating at symmetrically distributed
angles with respect to the coordinate axes. As shown below, this generalization
allows control over the distance between lattice sites.

The lattice potential for this arrangement can be written as\begin{equation}
V_{{\rm Latt}}(\mathbf{r})=\frac{1}{2}V_{0}\cos(\mathbf{b}_{1}\cdot\mathbf{r})+\frac{1}{2}V_{0}\cos(\mathbf{b}_{2}\cdot\mathbf{r}),\label{eq:V4B}\end{equation}
where\begin{eqnarray}
\mathbf{b}_{1}\equiv\mathbf{k}_{1}-\mathbf{k}_{2} & = & 2k\sin(\phi/2)\hat{\mathbf{x}},\label{eq:b14B}\\
\mathbf{b}_{2}\equiv\mathbf{k}_{3}-\mathbf{k}_{4} & = & 2k\sin(\phi/2)\hat{\mathbf{y}},\label{eq:b24B}\end{eqnarray}
are the reciprocal lattice vectors and $V_{0}=-{\rm Re}\{\alpha\} I_{0}/2\epsilon_{0}c$
is four times the dipole shift associated with a single laser of intensity
$I_{0}$. We will refer to $|V_{0}|$ as the \emph{light-shift strength},
and note that $|V_{0}|$ is the potential depth for the 1D lattice
produced by two intersecting light fields of equal intensity $I_{0}$.
In more than one spatial dimension the potential saddle point between
sites is the most essential characterization of the \emph{lattice
depth}, however this depends strongly on the lattice geometry and
is difficult to characterize. To be of most relevance to experiments,
in this paper we will compare different lattices made with the same
light-shift strength $V_{0}$, i.e. produced by sets of laser fields
with the same intensity. We note that $V_{0}$ can be a positive or
negative quantity depending on whether the light fields are blue or
red detuned from atomic resonance respectively. The 4 beam lattice
is symmetric with respect to $V_{0}$ changing sign (to within an
overall translation of the lattice), however the next lattice we consider
is not. 

The direct lattice vectors which specify the translation vectors between
lattices sites are \begin{eqnarray}
\mathbf{a}_{1} & = & \frac{\lambda}{2}\csc(\phi/2)\hat{\mathbf{x}},\label{eq:a14B}\\
\mathbf{a}_{2} & = & \frac{\lambda}{2}\csc(\phi/2)\hat{\mathbf{y}},\label{eq:a24B}\end{eqnarray}
 demonstrating that the lattice site spacing can be varied from $\lambda/2\to\infty$
as $\phi$ changes from $180^{\circ}\to0$.

\paragraph*{3 Beam Lattice}

The second 2D optical lattice we consider consists of 3 laser fields
--- the minimum number needed to generate a 2D lattice. We will refer
to this configuration as the 3 beam (3B) lattice. The geometric arrangement
of the light fields is shown in Fig. \ref{cap:LASERdiag}(b), with
all three light fields taken to have the same frequency. The lattice
potential, for this arrangement, given by\begin{eqnarray}
V_{{\rm {\rm Latt}}}(\mathbf{r}) & = & \frac{1}{2}V_{0}\cos(\mathbf{b}_{1}\cdot\mathbf{r})+\frac{1}{2}V_{0}\cos(\mathbf{b}_{2}\cdot\mathbf{r})\label{eq:V3B}\\
 &  & +\frac{1}{2}V_{0}\cos([\mathbf{b}_{1}+\mathbf{b}_{2}]\cdot\mathbf{r}),\nonumber \end{eqnarray}
is a sum of three 1D lattices arising from the interference between
each distinct pair of fields. As for the 4 beam case, we have introduced
the \emph{}light-shift strength parameter $V_{0}=-{\rm Re}\{\alpha\} I_{0}/2\epsilon_{0}c$,
and we define the reciprocal lattice vectors $\mathbf{b}_{j}$ below. 

The incident wave-vectors are of the form \begin{equation}
\mathbf{k}_{j}=k[\cos\theta_{j}\hat{\mathbf{x}}+\sin\theta_{j}\hat{\mathbf{y}}],\label{eq:kj3B}\end{equation}
 and generate a potential that exhibits a broad range of crystallographic
configurations, dependent on the choice of $\{\theta_{1},\theta_{2},\theta_{3}\}$.
The crystallography of this type of lattice has been considered by
Petsas \emph{et al.} \cite{Petsas1994a} in a context relevant to
laser cooling atoms, however we find it useful to cast their observations
in an analytic framework relevant to conservative lattices. To do
this we choose to label our wave vectors such that $\theta_{1}<\theta_{2}<\theta_{3}$,
and place the following restrictions on the angles: \begin{eqnarray}
 & \theta_{1}=0,\label{eq:angle_cond1}\\
0< & \theta_{2} & <180^{\circ},\label{eq:angle_cond2}\\
0< & \theta_{3}-\theta_{2} & <180^{\circ},\label{eq:angle_cond3}\end{eqnarray}
 For any set of three coplanar light fields (excluding the case where
two beams are co-propagating which gives rise to a 1D lattice) the
angles can always be transformed to satisfy requirements (\ref{eq:angle_cond1})-(\ref{eq:angle_cond3})
with an overall rotation the system in the $x$-$y$ plane.

Analogously to what was done in the four beam lattice, we define the
reciprocal lattice vectors as \begin{equation}
\mathbf{b}_{j}=\mathbf{k}_{j}-\mathbf{k}_{j+1}\quad(j=1,2).\label{eq:bj3B}\end{equation}
 We note that other choices of reciprocal lattice vectors are possible.
The direct lattice vectors $\{\mathbf{a}_{1},\mathbf{a}_{2}\}$ can
then be determined from the orthogonality relationship $\mathbf{a}_{i}\cdot\mathbf{b}_{j}=2\pi\delta_{ij}$
(e.g. see \cite{Mermin1976}). For the purposes of discussing direct
lattice geometry, it is useful to specify the angle $\phi_{D}$ between
the direct lattice vectors and the ratio of their lengths $r_{D}$.
It can be shown that these relate to the incident laser angles as
\begin{eqnarray}
\phi_{D} & = & 180^{\circ}-\frac{\theta_{3}}{2},\label{eq:phiD3B}\\
r_{D}\equiv\frac{|\mathbf{a}_{2}|}{|\mathbf{a}_{1}|} & = & \frac{\sin(\theta_{2}/2)}{\sin([\theta_{3}-\theta_{2}]/2)}.\label{eq:rD3B}\end{eqnarray}
This expression can be inverted to uniquely determine the laser angles
under restrictions (\ref{eq:angle_cond1})-(\ref{eq:angle_cond3})
for $r_{D}\in(0,\infty),$ and $\phi_{D}\in(\cos^{-1}(1/r_{D}),180^{\circ})$.

Finally, we note that unlike the 4 beam potential, the 3 beam potential
(\ref{eq:V3B}) is not symmetric with respect to sign change of $V_{0}$
(i.e. change in the sign of the laser detuning from atomic resonance),
a point we elaborate on in Sec. \ref{sec:Lattice-Geometry}. The 3
beam lattice only realizes a periodic set of confining wells for the
case of red detuning, i.e. $V_{0}<0$, and as such we restrict our
attention to red-detuned lattices in this paper.

\subsection{Wannier states \label{sub:Wannier-states}}

The eigenstates of the periodic single particle Hamiltonian, \begin{equation}
\hat{H}_{0}=\hat{\mathbf{p}}^{2}/2m+V_{{\rm Latt}}(\mathbf{r}),\label{EQN_H0}\end{equation}
 are known as Bloch states, which we write as $\psi_{n\mathbf{q}}(\mathbf{r})$,
with respective energy eigenvalue $\hbar\omega_{n\mathbf{q}}$. Here
we take the lattice potential to be a general periodic function on
the Bravais lattice of points $\mathbf{R}_{j}=n_{1j}\mathbf{a}_{1}+n_{2j}\mathbf{a}_{2}$
$(n_{ij}\in\mathbb{Z}$), such that $V_{{\rm Latt}}(\mathbf{r)=V_{{\rm Latt}}(r}+\mathbf{R}_{j})$.
According to Bloch's theorem, these eigenstates can be written as\begin{equation}
\psi_{n\mathbf{q}}(\mathbf{r})=u_{n\mathbf{q}}(\mathbf{r})e^{i\mathbf{q}\cdot\mathbf{r}},\end{equation}
 where $u_{n\mathbf{q}}(\mathbf{r})=u_{n\mathbf{q}}(\mathbf{r}+\mathbf{a}_{i})$
is a periodic function, $\mathbf{q}$ is the quasi-momentum, and $n$
is the band index. Formally, a Bravais lattice is of infinite extent,
however it is numerically convenient for us to consider a finite lattice
of $N$ sites with periodic boundary conditions. For this situation
each energy band contains $N$ eigenstates, and the quasi-momenta
can be restricted to the first Brillouin zone.

Bloch states are extended states that span the entire lattice. To
consider many-body effects arising from the interactions between particles
it is more convenient to work with a set of localized states. Here
we consider the Wannier basis which is a unitary transformation of
the Bloch basis, where the basis states are labeled by lattice site
position. The Wannier state centered at site $\mathbf{R}_{i}$ is
defined as \begin{equation}
w_{n}(\mathbf{r}-\mathbf{R}_{i})\equiv\frac{1}{\sqrt{N}}\sum_{\mathbf{q}}e^{-i\mathbf{q}\cdot\mathbf{R}_{i}}\psi_{n\mathbf{q}}(\mathbf{r}),\label{eq:WannDefn}\end{equation}
 where the summation is taken over all Bloch states of band-$n$.
For describing quantum degenerate Bose-gases in optical lattices we
will only consider Wannier states of the ground band. 

Wannier states are not eigenstates of the Hamiltonian and so an atom
prepared into a Wannier state at a given site will tunnel to other
Wannier states over time. This tunneling between sites is characterized
by the matrix elements of the Hamiltonian between the respective Wannier
states at those sites, i.e. \begin{eqnarray}
\gamma_{\mathbf{R}_{i}-\mathbf{R}_{j}} & \equiv & \int d\mathbf{r}\, w_{0}^{*}(\mathbf{r}-\mathbf{R}_{i})\hat{H}_{0}w_{0}(\mathbf{r}-\mathbf{R}_{j}),\label{EqnLattHamWan}\end{eqnarray}
 which reduces to\begin{equation}
\gamma_{\mathbf{R}_{i}-\mathbf{R}_{j}}\equiv\frac{1}{N}\sum_{\mathbf{q}}\hbar\omega_{n\mathbf{q}}e^{-i\mathbf{q}\cdot(\mathbf{R}_{j}-\mathbf{R}_{i})},\label{eq:tunneling}\end{equation}
 i.e. the Fourier transform of the Bloch dispersion relation. We note
that the overlap integral between states at different sites (\ref{eq:tunneling})
only depends on the relative separation between the sites, so that
characterizing the tunneling properties of a single Wannier state
completely determines the matrix elements of the whole lattice. 

For deep lattices the Wannier state is well localized at its central
site and $\gamma_{\mathbf{R}}$ rapidly decreases with increasing
relative separation $|\mathbf{R}|$. This is the well-known tight
binding limit, where a state at any given site only couples (to any
degree of significance) to a few neighboring sites. In this limit
only the most prominent tunneling matrix elements to other sites are
retained, where these sites are referred to as the \emph{nearest neighbors}.
For cases where the nearest neighbor tunneling terms are all equal
it is conventional to define the tunneling parameter or hopping rate
as $J=-\gamma_{\mathbf{R_{j}}}$ (e.g. see \cite{Jaksch1998a}). We
will present quantitative results for these tunneling rates and demonstrate
systems which exhibit different numbers of nearest neighbors in Sec.
\ref{sec:Wannier-States}.

\subsection{Bose-Hubbard model}

\label{SECBHmodel} Here we review the Bose-Hubbard model for interacting
Bosonic atoms in an optical lattice, closely following the original
derivation by Jaksch \emph{et al.} \cite{Jaksch1998a}. The usual
starting point for discussing many-body aspects of ultra-cold atoms
is the second quantized Hamiltonian \begin{eqnarray}
\hat{H} & = & \int d\mathbf{r}\,\hat{\psi}^{\dagger}(\mathbf{r})\left\{ \hat{H}_{0}+V_{{\rm ext}}(\mathbf{r})\right\} \hat{\psi}(\mathbf{r})\nonumber \\
 &  & +\frac{1}{2}g\int d\mathbf{r}\,\hat{\psi}^{\dagger}(\mathbf{r})\hat{\psi}^{\dagger}(\mathbf{r})\hat{\psi}(\mathbf{r})\hat{\psi}(\mathbf{r}),\label{EqnfullHamiltonian}\end{eqnarray}
 where $\hat{\psi}(\mathbf{r})$ is the boson field operator, $V_{{\rm ext}}(\mathbf{r})$
describes a slowly varying external potential (in addition to the
lattice potential in $\hat{H}_{0}$), $g=4\pi\hbar^{2}a_{s}/m$ is
the interaction strength, with $a_{s}$ the scattering length and
$m$ the atomic mass.

If a sample of sufficiently cold Bosonic atoms is loaded into a deep
enough lattice (e.g. from a Bose-Einstein condensate), then only states
of the lowest vibrational band will be populated. This system is conveniently
described by expanding the field operators in the Wannier basis of
the ground band \begin{equation}
\hat{\psi}(\mathbf{r})=\sum_{i}w_{0}(\mathbf{r}-\mathbf{R}_{i})\hat{a}_{i},\end{equation}
 where the operator $\hat{a}_{i}$ destroys a particle in the $\mathbf{R}_{i}$-Wannier
orbital, and satisfies the usual Bose commutation relations $[\hat{a}_{i},\hat{a}_{j}]=0$
and $[\hat{a}_{i},\hat{a}_{j}^{\dagger}]=\delta_{ij}$. In the Wannier
representation the Hamiltonian (\ref{EqnfullHamiltonian}) reduces
to the Bose-Hubbard form \begin{eqnarray}
\hat{H} & \approx & -J\sum_{\langle i,j\rangle}\hat{a}_{i}^{\dagger}\hat{a}_{j}+\sum_{i}\epsilon_{i}\hat{a}_{i}^{\dagger}\hat{a}_{i}\nonumber \\
 &  & +\frac{1}{2}U_{{\rm BH}}\sum_{i}\hat{a}_{i}^{\dagger}\hat{a}_{i}^{\dagger}\hat{a}_{i}\hat{a}_{i},\label{EqnTBHamiltonian}\end{eqnarray}
 where the first summation in Eq. (\ref{EqnfullHamiltonian}) has
been restricted to the nearest neighbor sites (i.e. tight binding
approximation discussed above). We have also introduced the site dependent
local energy \begin{eqnarray}
\epsilon_{i} & \equiv & \int d\mathbf{r}\, w_{0}^{*}(\mathbf{r}-\mathbf{R}_{i})V_{{\rm ext}}(\mathbf{r})w_{0}(\mathbf{r}-\mathbf{R}_{i}),\\
 & \simeq & V_{{\rm ext}}(\mathbf{R}_{i})\end{eqnarray}
 and the on-site interaction term \begin{equation}
U_{{\rm BH}}\equiv g\int d\mathbf{r}\,|w_{0}(\mathbf{r})|^{4}.\label{EQN_UBH}\end{equation}

The Bose-Hubbard model will be a good description of the system if
degrees of freedom in the higher bands can be neglected. For an energy
gap to the first excited band of $E_{{\rm gap}}$, we find the validity
conditions to be $k_{B}T\ll E_{{\rm gap}}$ and $nU_{{\rm BH}}\ll E_{{\rm gap}}$,
where $n$ is the average site occupation. Typically these requirements
are well satisfied in experiments. 

In this paper we calculate 2D Bose-Hubbard model parameters that can
be achieved using a 2D optical lattice. The nature of the confinement
transverse to the lattice will have considerable influence on the
physical properties of the system, however this is beyond the scope
of this paper. For simplicity we assume that the motion transverse
to the lattice is frozen out, e.g. as could be arranged experimentally
by superimposing a deep 1D lattice to slice the potential tubes of
the 2D lattice into a set of independent planes.

\subsection{Superfluid to Mott-insulator transition\label{sub:Superfluid-to-Mott-insulator}}

One of the most fascinating properties of the Bose-Hubbard Hamiltonian
is that it predicts a quantum-phase transition of the ground state
of the system between superfluid and Mott-insulating states \cite{Fisher1989a}.
The phase diagram for the system depends precisely on the Bose-Hubbard
parameters and the average filling factor of the system. It is convenient
to define the dimensionless quantity\begin{equation}
\eta\equiv\frac{U_{{\rm BH}}}{zJ},\label{eq:eta}\end{equation}
 embodying the Bose-Hubbard model parameters, where $z$ is the number
of nearest neighbors. For a homogeneous system with unit filling (i.e.
an average density of one boson per site), a critical value of $\eta_{c}\approx5.8$
is predicted by mean-field theories for the transition %
\footnote{E.g. see \cite{Jaksch1998a} and references therein.%
}. For $\eta<\eta_{c}$ the system is in a superfluid state where the
hopping between sites dominates over on-site repulsion. In this state,
the system exhibits phase coherence across the lattice and significant
fluctuations in the number of bosons per site. For $\eta>\eta_{c}$
interaction effects dominate tunneling and the particles become localized
--- this is the Mott-insulator state, characterized by a near definite
number of particles per site and an absence of phase coherence between
sites.

For bosonic atoms in an optical lattice the values of $U_{{\rm BH}}$
and $J$ can be dynamically modified by changing the lattice depth.
E.g. the effect of increasing the lattice depth is to confine the
bosons more tightly at each site, increasing $U_{{\rm BH}}$, and
reducing the tunneling between sites. Greiner \emph{et al.} \cite{Greiner2002a}
\textbf{}have experimentally demonstrated a loading procedure that
is sufficiently adiabatic for the system to be reversibly taken through
the phase transition.

\section{Lattice Geometry\label{sec:Lattice-Geometry}}

\begin{widetext}

\begin{figure}[!htb]
\includegraphics{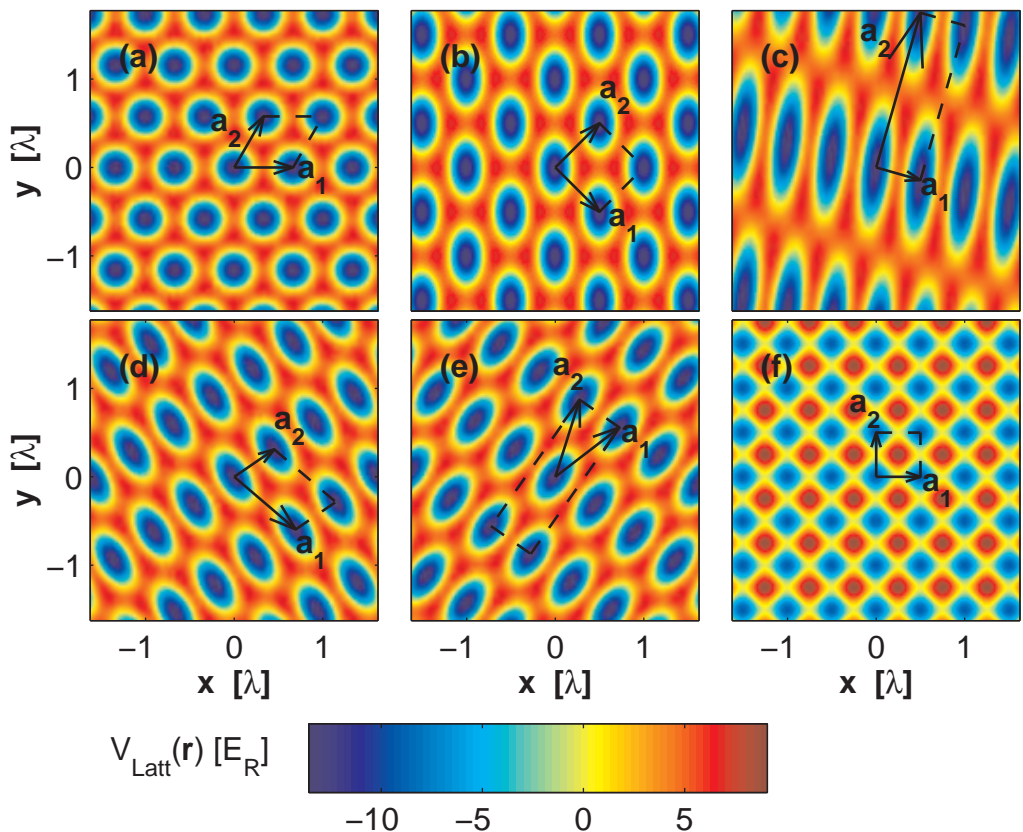}
\caption{\label{cap:FIG LattGeom} Optical lattice potentials. (a)-(e) Potentials
for the 3 beam laser configuration. (f) Potential for the 4 beam counter
propagating configuration. For these plots $V_{0}=-9E_{R}$, other
parameters discussed in the text. }
\end{figure}

\end{widetext}

In Fig. \ref{cap:FIG LattGeom} we demonstrate a variety of the possible
2D optical lattice geometries that can be made with the lattice types
discussed in Sec. \ref{sub:Optical-Lattice}. In particular, the 3
beam configuration allows the five distinct types of two-dimensional
Bravais lattices to be realized by suitably choosing the incident
light field angles (Figs. \ref{cap:FIG LattGeom}(a)-(e)). In Fig.
\ref{cap:FIG LattGeom}(f) we show the counter propagating 4 beam
lattice potential. Note that we have chosen to specify light-shift
strength in units of photon recoil energy, defined as $E_{R}=\hbar^{2}k^{2}/2m.$

In more detail the cases considered in Fig. \ref{cap:FIG LattGeom}
are (named according to the underlying Bravais lattice geometry):

\begin{itemize}
\item (a) \textbf{{[}3 Beam{]}} \textbf{Hexagonal Lattice:} This highly
symmetric triangular lattice displays hexad symmetry (symmetry under
$60^{\circ}$ rotations). This lattice naturally arises in close packing
of spheres - and corresponds to the planes of three dimensional hexagonal
close-packed and face-centered cubic lattices \cite{Mermin1976}.
The direct lattice vectors have length ratio $r_{D}=1$ and intersect
at an angle of $\phi_{D}=60^{\circ}$. The light fields needed to
make this lattice have planar angles of $\theta_{2}=120.0^{\circ}$
and $\theta_{3}=240.0^{\circ}$ (see Eqs. (\ref{eq:phiD3B}) and (\ref{eq:rD3B})). 
\item (b) \textbf{{[}3 Beam{]} Square Lattice:} This square lattice has
an underlying Bravais lattice with tetrad symmetry (symmetric under
$90^{\circ}$ rotations), though the potential does not exhibit this
symmetry (it is only symmetric under $180^{\circ}$ rotations). The
direct lattice vectors have length ratio $r_{D}=1$ and intersect
at an angle of $\phi_{D}=90^{\circ}$. The light fields needed to
make this lattice have planar angles of $\theta_{2}=90.0^{\circ}$
and $\theta_{3}=180.0^{\circ}$. 
\item (c) \textbf{{[}3 Beam{]}} \textbf{Rectangular Lattice:} The rectangular
lattice emphasizes the considerable asymmetry between directions that
can be engineered. For the specific case we have chosen to consider
the direct lattice vectors have length ratio $r_{D}=3.5$ and intersect
at an angle of $\phi_{D}=90^{\circ}$. The large lattice vector $\mathbf{a}_{2}$
arises because the light fields with wavevector $\mathbf{k}_{2}$
and $\mathbf{k}_{3}$ differ by a relatively small angle of $31.9^{\circ}$,
i.e. $\theta_{2}=148.1^{\circ}$ and $\theta_{3}=180.0^{\circ}$. 
\item (d) \textbf{{[}3 Beam{]} Oblique Lattice:} This is the most general
planar lattice type of which all other 2D Bravais lattices are special
cases. The only symmetry that this lattice exhibits in general is
diad symmetry (symmetric under $180^{\circ}$ rotations). For the
specific case considered, the direct lattice vectors have length ratio
$r_{D}=0.6$ and intersect at an angle of $\phi_{D}=85^{\circ}$.
The light fields needed to make this lattice have planar angles of
$\theta_{2}=64.5^{\circ}$ and $\theta_{3}=190.0^{\circ}$. 
\item (e) \textbf{{[}3 Beam{]} Centered Rectangular Lattice:} The centered
rectangular lattice is usually specified using the larger (non-primitive)
cell indicated in Fig. \ref{cap:FIG LattGeom}(e), that emphasizes
the presence of a face centered point. For the specific case considered,
the direct lattice vectors have length ratio $r_{D}=1$ and intersect
at an angle of $\phi_{D}=35^{\circ}$. The light fields needed to
make this lattice have planar angles of $\theta_{2}=145.0^{\circ}$
and $\theta_{3}=290.0^{\circ}$. 
\item (f) \textbf{{[}4 Beam{]}} \textbf{Square Lattice:} This lattice is
formed from two sets of counter propagating light fields with a $\lambda/2$
lattice site spacing, and exhibits the highest lattice site density
(per unit area) of the lattices considered in Fig. \ref{cap:FIG LattGeom}. 
\end{itemize}
Before considering the Wannier states of these lattices it is useful
to compare the most significant differences between the 3 beam and
4 beam lattices.

\paragraph*{Lattice geometry: }

As demonstrated in Figs. \ref{cap:FIG LattGeom}(a)-(e), the 3 beam
lattice exhibits a wide range of lattice types, with the laser angles
calculated using expressions (\ref{eq:phiD3B}) and (\ref{eq:rD3B}).
The rigid constraints we placed upon the light field arrangement of
the 4 beam lattice ensures the resulting lattice is square. Relaxing
these constraints by allowing the relative intersection angle of each
pair to be different, or apply an overall rotation to one pair, it
is possible to generate rectangular or oblique lattices respectively.
These generalizations are not considered in this work. 

The lattice geometry plays a central role in determining the number
of nearest neighbors, $z$, in the derived Bose-Hubbard model. Inspection
of Fig. \ref{cap:FIG LattGeom} suggests that the 3 beam lattice offers
cases that realize 2 (see Fig. \ref{cap:FIG LattGeom}(c)), 4 (see
Fig. \ref{cap:FIG LattGeom}(b)), and 6 (see Fig. \ref{cap:FIG LattGeom}(a))
nearest neighbors. With the aide of band structure calculations for
the tunneling matrix elements we return to this issue in the next
section, and verify the actual numbers of nearest neighbors. Another
important factor relating to the geometry is the distance between
lattice sites. For the 4 beam lattice (Fig. \ref{cap:FIG LattGeom}(f))
the spacing between sites is the minimum value of $|\mathbf{a}_{j}|=\lambda/2$.
For the 3 beam lattice spacing is generally larger, e.g. $|\mathbf{a}_{j}|=2\lambda/3$
for the hexagonal lattice (Fig. \ref{cap:FIG LattGeom}(a)), $|\mathbf{a}_{j}|=\lambda/\sqrt{2}$
in the square lattice (Fig. \ref{cap:FIG LattGeom}(b)). The site
spacing will influence the tunneling between sites, as this is strongly
dependent on the \emph{barrier width} between sites.

\paragraph*{Unit cell potential shape: }

\begin{figure}[!htb]
\includegraphics[%
  width=3.4in,
  keepaspectratio]{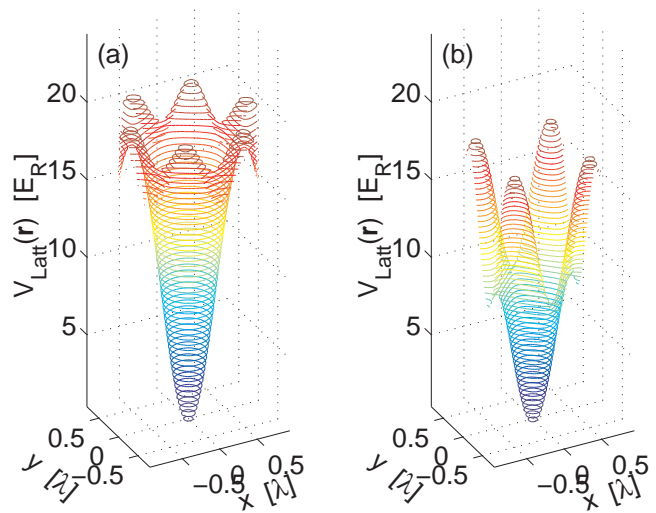}

\caption{\label{cap:FIGpotsites} Comparison of the individual potential wells
of the 3 beam and 4 beam lattices. Potential energy contour plot of
(a) 3 beam hexagonal lattice, and (b) 4 beam square lattice. The potentials
are displayed on a region of space slightly larger than a primitive
unit cell. The lattice potentials displayed here correspond to the
same configurations and parameters used in Figs. \ref{cap:FIG LattGeom}(a)
and (f) respectively. An energy offset has been added so that the
potential minima are at 0. }
\end{figure}

The single well potential for the 3 beam and 4 beam lattices are quite
distinctive, as is illustrated for two particular cases in Fig. \ref{cap:FIGpotsites}.
The most noticeable feature distinguishing these cases (and is a general
observation for the different configurations of 3 beam lattice) is
that the saddle point of the potential separating a given lattice
site from the neighboring site is much higher in the 3 beam lattice
(Fig. \ref{cap:FIGpotsites}(a)) than in the 4 beam lattice (Fig.
\ref{cap:FIGpotsites}(b)). The superior confinement of the 3 beam
lattice produces much better localized Wannier states than for a 4
beam lattice with the same lattice site spacing and field intensity
(per incident beam), as we demonstrate in the next section. From Fig.
\ref{cap:FIGpotsites}(a) it is clear why the 3 beam lattice is not
symmetric with respect to $V_{0}$ changing sign: The subtle crown-like
maxima of the 3 beam lattice (which become the minima when the light
fields are changed to blue detuning) are shallow and provide little
confinement. The dominant feature for the blue detuned 3 beam lattice
is the periodic array of maxima, referred to as an \emph{anti-dot}
lattice (e.g. see \cite{Petsas1999a}).

\section{Wannier States and Bose-Hubbard Parameters\label{sec:Wannier-States}}

\begin{widetext}

\begin{figure}[!htb]
\includegraphics{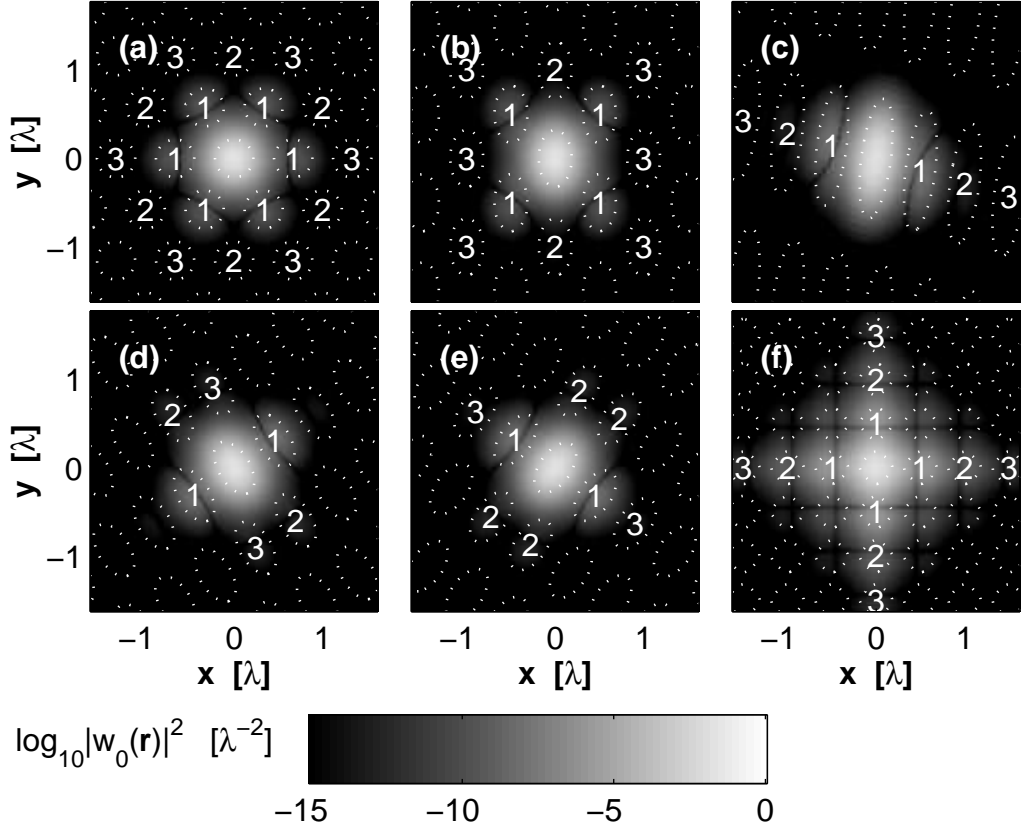}

\caption{\label{cap:FIG Wannier} Wannier state density distributions and
tunneling neighborhoods. The $\log$-densities of the Wannier states
in (a)-(f) correspond to the similarly labeled potentials displayed
in Fig. \ref{cap:FIG LattGeom}. Potential energy contours are indicated
by dotted white lines mark the lattice site locations. The white numbers
centered upon neighboring sites denote the 'nearness' of that site
according to a ranking of the tunneling matrix elements (see text).
For these plots $V_{0}=-9E_{R}$.}
\end{figure}

\end{widetext}

In this section we present numerical calculations of the Wannier states
and Bose-Hubbard parameters for the different 2D optical lattices
considered in this paper. 

To begin, we briefly summarize our numerical procedures. We diagonalize
the non-interacting lattice Hamiltonian (\ref{EQN_H0}) represented
in a plane-wave expansion to obtain the ground band Bloch states and
energy eigenvalues for each quasi-momentum value within the first
Brillouin zone. Ensuring the Bloch states are continuous in quasi-momentum,
we construct the Wannier states by numerically summing the Bloch states
with the appropriate phase factors (see Eq. (\ref{eq:WannDefn})).
Having obtained the Wannier states, the on-site interaction can be
evaluated according to Eq. (\ref{EQN_UBH}) using numerical integration.
Finally, the tunneling matrix elements are calculated from the ground
band energy spectrum using Eq. (\ref{eq:tunneling}). We note that
to numerically calculate the tunneling matrix elements couplings,
$\gamma_{\mathbf{R}_{i}-\mathbf{R}_{j}}$, we need to take the lattice
size $N$ to be sufficiently large to contain the sites $\mathbf{R}_{i}$
and $\mathbf{R}_{j}$. 

The Wannier state localized at the origin site for each of the six
lattice potentials of Fig. \ref{cap:FIG LattGeom} are shown in Fig.
\ref{cap:FIG Wannier}. At the light-shift strength considered ($V_{0}=-9E_{R}$),
the Wannier states are all well localized upon the central site, and
to facilitate observing the small density tails at the neighboring
sites we have employed a $\log$-scale for the density color-map.
The Wannier state of the 4 beam square lattice is the least localized
of the all the states considered, mainly due to poor degree of lattice
site confinement discussed earlier (see Fig. \ref{cap:FIGpotsites}(b)).
It is interesting to note that the first node in the Wannier states
occurs near the middle of the neighboring site, as is required for
states at different sites to be orthogonal.

The tunneling matrix elements provide a quantitative measure of the
coupling strength between different Wannier states. In order to demonstrate
the variable number of nearest neighbors, and the extent to which
further than nearest neighbor couplings are negligible, we have listed
the tunneling matrix elements in table \ref{cap:WanTable} and have
indicated the groups of sites these matrix elements correspond to
in Fig. \ref{cap:FIG Wannier}. To be clear, in Fig. \ref{cap:FIG Wannier}
we have enumerated all sites that have the same tunneling matrix element
to the central site with the same number, e.g. 1 indicates 'first
nearest neighbor', 2 indicates '2nd nearest neighbor', etc. (i.e.
larger integers indicate decreasing magnitude of tunneling matrix
element). These results indicate that at $V_{0}=-9E_{R}$ the second
nearest neighbors have tunneling rates that are 2-4 orders of magnitude
smaller than those of the nearest neighbors. Tight-binding can be
expected to provide an excellent description of the system in this
regime, i.e. we are justified in identifying the first column of Table
\ref{cap:WanTable} with $-J$ (see Sec. \ref{sub:Wannier-states}),
and ignoring all other couplings. Another important observation is
that the number of nearest neighbors for the Wannier states in Figs.
\ref{cap:FIG Wannier}(a)-(f) (i.e. the number of $1$'s in each figure)
takes the values 2, 4, or 6. 

\begin{table}[!htb]
\begin{tabular}{|l|c|c|c|}
\hline 
$\qquad$site label:&
1-sites&
2-sites&
3-sites\tabularnewline
\hline
\hline 
(a) (3B) Hexagonal&
$-2.79\times10^{-4}$&
$8.75\times10^{-8}$&
$5.64\times10^{-8}$\tabularnewline
\hline 
(b) (3B) Square&
$-1.862\times10^{-4}$&
$-1.12\times10^{-6}$&
$-8.88\times10^{-8}$\tabularnewline
\hline 
(c) (3B) Rectangular&
$-2.90\times10^{-3}$&
$4.54\times10^{-6}$&
$-1.12\times10^{-8}$\tabularnewline
\hline 
(d) (3B) Oblique&
$-1.83\times10^{-3}$&
$-6.48\times10^{-6}$&
$-4.28\times10^{-6}$\tabularnewline
\hline 
(e) (3B) Centered&
$-1.80\times10^{-3}$&
$-6.33\times10^{-6}$&
$1.84\times10^{-6}$\tabularnewline
\hline 
(f) (4B) Square&
$-2.42\times10^{-2}$&
$3.77\times10^{-4}$&
$-9.28\times10^{-6}$\tabularnewline
\hline
\end{tabular}

\caption{\label{cap:WanTable} Values of the first three nearest neighbor
couplings for the Wannier states considered in Fig. \ref{cap:FIG Wannier}
in units of $E_{R}$. The values in the columns labeled {}``$j$-sites''
correspond to the tunneling matrix element between the Wannier state
at the origin and the sites enumerated by $j$ in Fig. \ref{cap:FIG Wannier}.}
\end{table}
We note that the tunneling to the rows of sites vertically separated
from the site at the origin in Fig. \ref{cap:FIG Wannier}(c) is so
small that it is completely negligible compared to the first three
neighbors considered. Also, the separable nature of the potential
for the square 4 beam lattice in Fig. (f) means that tunneling to
all sites not lying on axis are identically zero.

As a final comparison between the lattices we consider the global
variation of the nearest neighbor tunneling and the on-site interaction
for a broad range of light-shift strengths. We have chosen five lattices
to make this comparison. The first four lattices are the most unique
of those considered in Figs. \ref{cap:FIG LattGeom} and \ref{cap:FIG Wannier}:
(a) the 3 beam hexagonal lattice (6 nearest neighbors); (b) the 3
beam square lattice (4 nearest neighbors); (c) the 3 beam rectangular
lattice (2 nearest neighbors); and (f) the 4 beam square lattice (4
nearest neighbors). For the fifth lattice we introduce a new 4 beam
configuration with a larger lattice constant of $|\mathbf{a}_{j}|=2\lambda/3$
(for which the pairs of light fields intersect at $\phi\approx97.2^{\circ}$),
chosen to have the same distance between sites as the hexagonal lattice.
Our motivation for considering this lattice is to better assess the
extent to which wider spacing between sites in the 4 beam lattice
reduces nearest neighbor tunneling. 

\begin{widetext}

\begin{figure}[!htb]
\includegraphics[%
  width=6in,
  keepaspectratio]{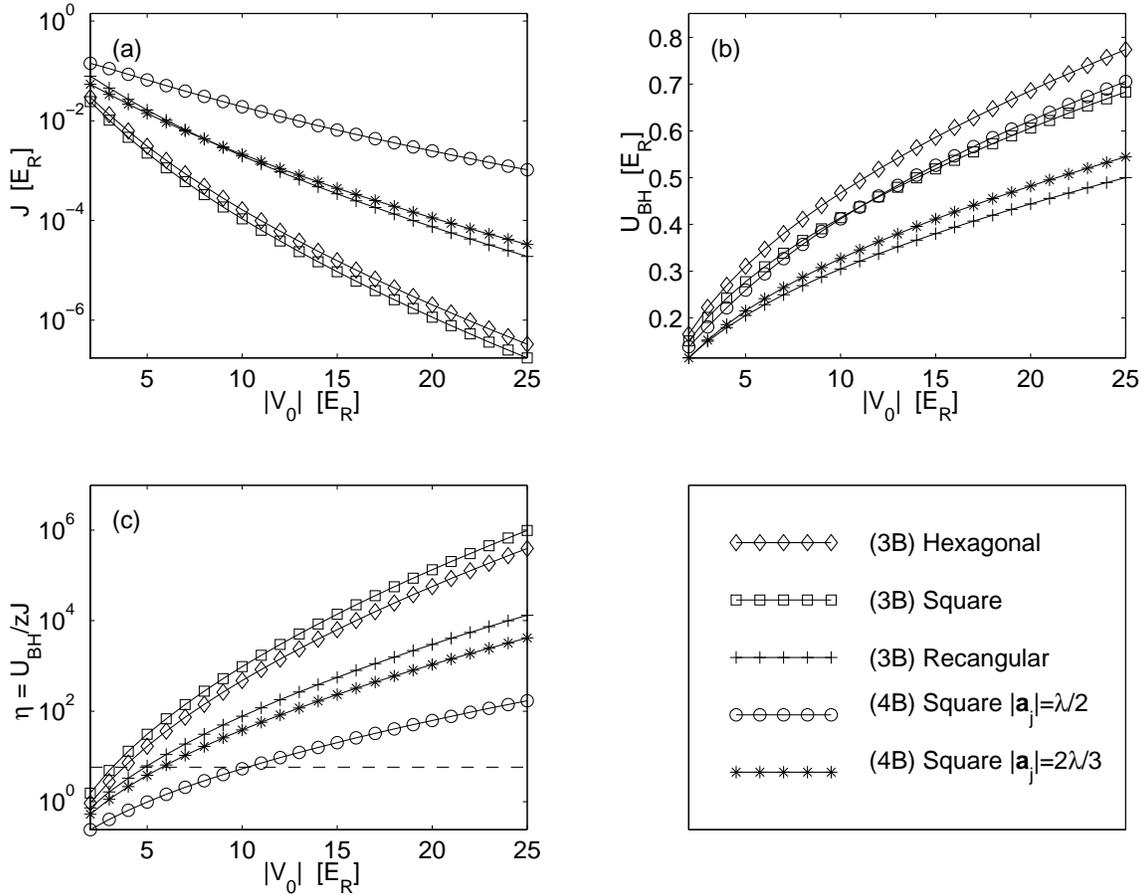}

\caption{\label{cap:FIG WannierUJ} Bose-Hubbard parameters for a variety
of different 2D lattices at various light-shift strengths (a) Nearest
neighbor tunneling strength. (b) On-site interaction strength. (c)
Dimensionless scaling parameter $\eta$ (see Sec. \ref{sub:Superfluid-to-Mott-insulator}).
Dashed horizontal line indicates the critical value for superfluid
to Mott-insulator transition for a homogeneous system with filling
factor $n=1$. We have taken an effective 2D coupling constant of
$g_{2D}=0.98\, E_{R}/k^{2}$ ($k=2\pi/\lambda$). Note that the coupling
strength $g$, discussed in Sec. \ref{SECBHmodel}, involves the 3D
scattering length, so that $g_{2D}$, is obtained by integrating over
the degree of freedom transverse to the lattice. The value of $g_{2D}$
we use here assumes the potential transverse to the lattice is a tightly
confining harmonic trap of frequency $\omega_{z}=hk^{2}/m$. We choose
physical parameters corresponding to the scattering length of $^{87}$Rb
and $\lambda=852$nm (similar to those of Ref. \cite{Greiner2002a}).}
\end{figure}

\end{widetext}

The results for the five chosen lattice types are shown in Fig. \ref{cap:FIG WannierUJ}.
In Fig. \ref{cap:FIG WannierUJ}(a) the nearest neighbor tunneling
matrix elements are displayed. Generally the 3 beam lattices have
comparable or considerably smaller tunneling matrix elements than
the 4 beam lattices at any given light-shift strength. In particular,
tunneling in the 3 beam square and hexagonal lattices is more that
3 orders of magnitude smaller than the counter propagating 4 beam
lattice at the maximum light-shift. Of the 4 beam lattices, the non-counter
propagating configuration has significantly smaller tunneling matrix
element than the counter propagating configuration at each light-shift
strength. Nevertheless, the non-counter propagating 4 beam lattice
(with lattice site spacing matching that of the hexagon lattice) has
much larger tunneling matrix elements than the hexagonal lattice due
to the residual differences in the potentials of the individual wells
(see Sec. \ref{sec:Lattice-Geometry}). 

In Fig. \ref{cap:FIG WannierUJ}(b) we show the on-site interaction
strength, $U_{{\rm BH}}$. It is surprising to note that for the hexagonal
lattice $U_{{\rm BH}}$ is higher than that of the counter propagating
4 beam lattice at every value of $V_{0}$, indicating that the hexagonal
lattice has tighter lattice site confinement. A useful measure of
the confinement is furnished by the oscillation frequency \emph{}of
the harmonic approximation to the lattice site minima. For fixed $V_{0}$,
we find that the oscillation frequency of the hexagonal lattice is
$\sim6$\% higher than for the counter-propagating 4 beam lattice. 

In Fig. \ref{cap:FIG WannierUJ}(c) we plot $\eta$ (see Eq. (\ref{eq:eta})),
which embodies the parameters of the Bose-Hubbard Hamiltonian. As
discussed in Sec. \ref{sub:Superfluid-to-Mott-insulator}, the value
of $\eta$ determines whether the equilibrium ground state is a superfluid
or a Mott-insulator. For reference we have indicated the transition
point for filling factor $n=1$ as a horizontal dashed line in Fig.
\ref{cap:FIG WannierUJ}(c). For the lattices considered, the Mott-transition
(where the curves cross the dashed line) varies over a wide range
of light-shift strengths. In the hexagonal lattice the Mott-transition
is at the lowest light-shift of $|V_{0}|\approx3.7E_{R}$, whereas
the transition for the 4 beam counter propagating lattice occurs at
the largest value of $|V_{0}|\approx10.2E_{R}$. For reference, this
4 beam lattice is the two dimensional version of the lattice recently
used by Greiner \emph{et al.} \cite{Greiner2002a} to experimentally
observe the Mott-transition. 

Comparing two 4 beam lattices in Figs. \ref{cap:FIG WannierUJ}(a)-(c)
yields the following observations. Relative to the $|\mathbf{a}_{j}|=\lambda/2$
lattice, the $|\mathbf{a}_{j}|=2\lambda/3$ lattice has significantly
smaller $J$, smaller $U_{{\rm BH}}$, which combined result in a
substantially larger $\eta$ value at each light-shift strength. In
general choosing smaller angles between the pairs of light fields
$\phi$ in the 4 beam lattice (see Fig. \ref{cap:LASERdiag}) could
be used to maximize $\eta$, and reduce the light-shift strength at
which the Mott-transition occurs. E.g., for $\phi=60^{\circ}$ the
lattice site spacing is $\lambda$, and Mott-transition occurs at
$|V_{0}|\approx2.5E_{R}$ --- lower than any of the 3 beam curves.
However, reducing the transition light-shift strength is made at the
expense of $U_{{\rm BH}}$, the tightness of well site confinement,
and the energy gap to higher bands (hence the validity of the Bose-Hubbard
description). Balancing these trade offs will be an important consideration
for many practical applications.

\section{Conclusions}

In this paper we have investigated different types of 2D optical lattices
that might be used in experiments to produce highly correlated states,
such as the Mott-insulator state, or to engineer 1D potentials. Through
numerical investigation of the band structure and explicit calculation
of the Wannier states, we have evaluated the Bose-Hubbard parameters
over a wide regime. We have demonstrated lattices that realize Bose-Hubbard
models with 2, 4, or 6 nearest neighbors and significantly effect
the light-shift strength at which superfluid to Mott-insulator transition
occurs.

\section*{Acknowledgments}

P.B.B would like to thank J. V. Porto and members of the NIST Laser
Cooling Group for useful discussions. This work was supported by the
US Office of Naval Research, and the Advanced Research and Development
Activity.


\begin{thebibliography}{21}
\expandafter\ifx\csname natexlab\endcsname\relax\def\natexlab#1{#1}\fi
\expandafter\ifx\csname bibnamefont\endcsname\relax
  \def\bibnamefont#1{#1}\fi
\expandafter\ifx\csname bibfnamefont\endcsname\relax
  \def\bibfnamefont#1{#1}\fi
\expandafter\ifx\csname citenamefont\endcsname\relax
  \def\citenamefont#1{#1}\fi
\expandafter\ifx\csname url\endcsname\relax
  \def\url#1{\texttt{#1}}\fi
\expandafter\ifx\csname urlprefix\endcsname\relax\def\urlprefix{URL }\fi
\providecommand{\bibinfo}[2]{#2}
\providecommand{\eprint}[2][]{\url{#2}}

\bibitem[{\citenamefont{Deutsch and Jessen}(1998)}]{Deutsch1998a}
\bibinfo{author}{\bibfnamefont{I.}~\bibnamefont{Deutsch}} \bibnamefont{and}
  \bibinfo{author}{\bibfnamefont{P.}~\bibnamefont{Jessen}},
  \bibinfo{journal}{Phys. Rev. A} \textbf{\bibinfo{volume}{57}},
  \bibinfo{pages}{1972} (\bibinfo{year}{1998}).

\bibitem[{\citenamefont{Calarco et~al.}(2000)\citenamefont{Calarco, Briegel,
  Jaksch, Cirac, and Zoller}}]{Calarco2000a}
\bibinfo{author}{\bibfnamefont{T.}~\bibnamefont{Calarco}},
  \bibinfo{author}{\bibfnamefont{H.}~\bibnamefont{Briegel}},
  \bibinfo{author}{\bibfnamefont{D.}~\bibnamefont{Jaksch}},
  \bibinfo{author}{\bibfnamefont{J.~I.} \bibnamefont{Cirac}}, \bibnamefont{and}
  \bibinfo{author}{\bibfnamefont{P.}~\bibnamefont{Zoller}},
  \bibinfo{journal}{J. of Mod. Opt.} \textbf{\bibinfo{volume}{47}},
  \bibinfo{pages}{2137} (\bibinfo{year}{2000}).

\bibitem[{\citenamefont{Jaksch and Zoller}(2003)}]{Jaksch2003a}
\bibinfo{author}{\bibfnamefont{D.}~\bibnamefont{Jaksch}} \bibnamefont{and}
  \bibinfo{author}{\bibfnamefont{P.}~\bibnamefont{Zoller}},
  \bibinfo{journal}{New J. Phys.} \textbf{\bibinfo{volume}{5}},
  \bibinfo{pages}{56} (\bibinfo{year}{2003}).

\bibitem[{\citenamefont{Jaksch et~al.}(1998)\citenamefont{Jaksch, Bruder,
  Cirac, Gardiner, and Zoller}}]{Jaksch1998a}
\bibinfo{author}{\bibfnamefont{D.}~\bibnamefont{Jaksch}},
  \bibinfo{author}{\bibfnamefont{C.}~\bibnamefont{Bruder}},
  \bibinfo{author}{\bibfnamefont{J.~I.} \bibnamefont{Cirac}},
  \bibinfo{author}{\bibfnamefont{C.}~\bibnamefont{Gardiner}}, \bibnamefont{and}
  \bibinfo{author}{\bibfnamefont{P.}~\bibnamefont{Zoller}},
  \bibinfo{journal}{Phys. Rev. Lett.} \textbf{\bibinfo{volume}{81}},
  \bibinfo{pages}{3108} (\bibinfo{year}{1998}).

\bibitem[{\citenamefont{Hofstetter et~al.}(2002)\citenamefont{Hofstetter,
  Cirac, Zoller, Demler, and Lukin}}]{Hofstetter2002a}
\bibinfo{author}{\bibfnamefont{W.}~\bibnamefont{Hofstetter}},
  \bibinfo{author}{\bibfnamefont{J.}~\bibnamefont{Cirac}},
  \bibinfo{author}{\bibfnamefont{P.}~\bibnamefont{Zoller}},
  \bibinfo{author}{\bibfnamefont{E.}~\bibnamefont{Demler}}, \bibnamefont{and}
  \bibinfo{author}{\bibfnamefont{M.}~\bibnamefont{Lukin}},
  \bibinfo{journal}{Phys. Rev. Lett.} \textbf{\bibinfo{volume}{89}},
  \bibinfo{pages}{220407} (\bibinfo{year}{2002}).

\bibitem[{\citenamefont{Orzel et~al.}(2001)\citenamefont{Orzel, Tuchman,
  Fenselau, Yasuda, and Kasevich}}]{Orzel2001a}
\bibinfo{author}{\bibfnamefont{C.}~\bibnamefont{Orzel}},
  \bibinfo{author}{\bibfnamefont{A.~K.} \bibnamefont{Tuchman}},
  \bibinfo{author}{\bibfnamefont{M.~L.} \bibnamefont{Fenselau}},
  \bibinfo{author}{\bibfnamefont{M.}~\bibnamefont{Yasuda}}, \bibnamefont{and}
  \bibinfo{author}{\bibfnamefont{M.~A.} \bibnamefont{Kasevich}},
  \bibinfo{journal}{Science} \textbf{\bibinfo{volume}{23}},
  \bibinfo{pages}{2386} (\bibinfo{year}{2001}).

\bibitem[{\citenamefont{Greiner
  et~al.}(2002{\natexlab{a}})\citenamefont{Greiner, Mandel, Esslinger,
  H{\"a}nsch, and Bloch}}]{Greiner2002a}
\bibinfo{author}{\bibfnamefont{M.}~\bibnamefont{Greiner}},
  \bibinfo{author}{\bibfnamefont{O.}~\bibnamefont{Mandel}},
  \bibinfo{author}{\bibfnamefont{T.}~\bibnamefont{Esslinger}},
  \bibinfo{author}{\bibfnamefont{T.~W.} \bibnamefont{H{\"a}nsch}},
  \bibnamefont{and} \bibinfo{author}{\bibfnamefont{I.}~\bibnamefont{Bloch}},
  \bibinfo{journal}{Nature} \textbf{\bibinfo{volume}{415}}, \bibinfo{pages}{39}
  (\bibinfo{year}{2002}{\natexlab{a}}).

\bibitem[{\citenamefont{Greiner
  et~al.}(2002{\natexlab{b}})\citenamefont{Greiner, Mandel, H{\"a}nsch, and
  Bloch}}]{Greiner2002b}
\bibinfo{author}{\bibfnamefont{M.}~\bibnamefont{Greiner}},
  \bibinfo{author}{\bibfnamefont{O.}~\bibnamefont{Mandel}},
  \bibinfo{author}{\bibfnamefont{T.~W.} \bibnamefont{H{\"a}nsch}},
  \bibnamefont{and} \bibinfo{author}{\bibfnamefont{I.}~\bibnamefont{Bloch}},
  \bibinfo{journal}{Nature} \textbf{\bibinfo{volume}{419}}, \bibinfo{pages}{51}
  (\bibinfo{year}{2002}{\natexlab{b}}).

\bibitem[{\citenamefont{Mandel et~al.}(2003)\citenamefont{Mandel, Greiner,
  Widera, Rom, H{\"a}nsch, and Bloch}}]{Mandel2003a}
\bibinfo{author}{\bibfnamefont{O.}~\bibnamefont{Mandel}},
  \bibinfo{author}{\bibfnamefont{M.}~\bibnamefont{Greiner}},
  \bibinfo{author}{\bibfnamefont{A.}~\bibnamefont{Widera}},
  \bibinfo{author}{\bibfnamefont{T.}~\bibnamefont{Rom}},
  \bibinfo{author}{\bibfnamefont{T.}~\bibnamefont{H{\"a}nsch}},
  \bibnamefont{and} \bibinfo{author}{\bibfnamefont{I.}~\bibnamefont{Bloch}},
  \bibinfo{journal}{Phys. Rev. Lett.} \textbf{\bibinfo{volume}{91}},
  \bibinfo{pages}{010407} (\bibinfo{year}{2003}).

\bibitem[{\citenamefont{Fisher et~al.}(1989)\citenamefont{Fisher, Weichman,
  Grinstein, and Fisher}}]{Fisher1989a}
\bibinfo{author}{\bibfnamefont{M.~P.~A.} \bibnamefont{Fisher}},
  \bibinfo{author}{\bibfnamefont{P.~B.} \bibnamefont{Weichman}},
  \bibinfo{author}{\bibfnamefont{G.}~\bibnamefont{Grinstein}},
  \bibnamefont{and} \bibinfo{author}{\bibfnamefont{D.~S.}
  \bibnamefont{Fisher}}, \bibinfo{journal}{Phys. Rev. B}
  \textbf{\bibinfo{volume}{40}}, \bibinfo{pages}{546} (\bibinfo{year}{1989}).

\bibitem[{\citenamefont{Burger et~al.}(2002)\citenamefont{Burger, Cataliotti,
  Fort, Maddaloni, Minardi, and Inguscio}}]{Burger2002a}
\bibinfo{author}{\bibfnamefont{S.}~\bibnamefont{Burger}},
  \bibinfo{author}{\bibfnamefont{F.~S.} \bibnamefont{Cataliotti}},
  \bibinfo{author}{\bibfnamefont{C.}~\bibnamefont{Fort}},
  \bibinfo{author}{\bibfnamefont{P.}~\bibnamefont{Maddaloni}},
  \bibinfo{author}{\bibfnamefont{F.}~\bibnamefont{Minardi}}, \bibnamefont{and}
  \bibinfo{author}{\bibfnamefont{M.}~\bibnamefont{Inguscio}},
  \bibinfo{journal}{Europhys. Latt.} \textbf{\bibinfo{volume}{57}},
  \bibinfo{pages}{1} (\bibinfo{year}{2002}).

\bibitem[{\citenamefont{Laburthe{-T}olra
  et~al.}(cond-mat/0312003)\citenamefont{Laburthe{-T}olra, Porto, O'Hara, Huckans, Rolston,
  and Phillips}}]{NIST2003a}
\bibinfo{author}{\bibfnamefont{B.}~\bibnamefont{Laburthe{-T}olra}},
  \bibinfo{author}{\bibfnamefont{J.~V.} \bibnamefont{Porto}},
  \bibinfo{author}{\bibfnamefont{K.~M.} \bibnamefont{O'Hara}},
  \bibinfo{author}{\bibfnamefont{J.~H.} \bibnamefont{Huckans}},
  \bibinfo{author}{\bibfnamefont{S.~L.} \bibnamefont{Rolston}},
  \bibnamefont{and} \bibinfo{author}{\bibfnamefont{W.~D.}
  \bibnamefont{Phillips}} \bibinfo{journal}{cond-matt},
  \bibinfo{pages}{0312003} (\bibinfo{year}{2003}).

\bibitem[{\citenamefont{Moritz et~al.}(2003)\citenamefont{Moritz, Stöferle,
  Köhl, and Esslinger}}]{Moritz2003a}
\bibinfo{author}{\bibfnamefont{H.}~\bibnamefont{Moritz}},
  \bibinfo{author}{\bibfnamefont{T.}~\bibnamefont{Stöferle}},
  \bibinfo{author}{\bibfnamefont{M.}~\bibnamefont{Köhl}}, \bibnamefont{and}
  \bibinfo{author}{\bibfnamefont{T.}~\bibnamefont{Esslinger}},
  \bibinfo{journal}{Phys. Rev. Lett.} \textbf{\bibinfo{volume}{91}},
  \bibinfo{pages}{250402} (\bibinfo{year}{2003}).

\bibitem[{\citenamefont{Abramowitz and Stegun}(1964)}]{HandBkMathFns}
\bibinfo{author}{\bibfnamefont{M.}~\bibnamefont{Abramowitz}} \bibnamefont{and}
  \bibinfo{author}{\bibfnamefont{I.~A.} \bibnamefont{Stegun}},
  \emph{\bibinfo{title}{Handbook of mathematical functions}}
  (\bibinfo{publisher}{US Government Printing Office}, \bibinfo{year}{1964}).

\bibitem[{\citenamefont{Greiner et~al.}(2001)\citenamefont{Greiner, Bloch,
  Mandel, H{\"a}nsch, and Esslinger}}]{Greiner2001a}
\bibinfo{author}{\bibfnamefont{M.}~\bibnamefont{Greiner}},
  \bibinfo{author}{\bibfnamefont{I.}~\bibnamefont{Bloch}},
  \bibinfo{author}{\bibfnamefont{O.}~\bibnamefont{Mandel}},
  \bibinfo{author}{\bibfnamefont{T.~W.} \bibnamefont{H{\"a}nsch}},
  \bibnamefont{and}
  \bibinfo{author}{\bibfnamefont{T.}~\bibnamefont{Esslinger}},
  \bibinfo{journal}{Phys. Rev. Lett.} \textbf{\bibinfo{volume}{87}},
  \bibinfo{pages}{160405} (\bibinfo{year}{2001}).

\bibitem[{\citenamefont{Peil et~al.}(2003)\citenamefont{Peil, Porto, Tolra,
  Obrecht, King, Subbotin, Rolston, and Phillips}}]{Peil2003a}
\bibinfo{author}{\bibfnamefont{S.}~\bibnamefont{Peil}},
  \bibinfo{author}{\bibfnamefont{J.~V.} \bibnamefont{Porto}},
  \bibinfo{author}{\bibfnamefont{B.~L.} \bibnamefont{Tolra}},
  \bibinfo{author}{\bibfnamefont{J.~M.} \bibnamefont{Obrecht}},
  \bibinfo{author}{\bibfnamefont{B.~E.} \bibnamefont{King}},
  \bibinfo{author}{\bibfnamefont{M.}~\bibnamefont{Subbotin}},
  \bibinfo{author}{\bibfnamefont{S.~L.} \bibnamefont{Rolston}},
  \bibnamefont{and} \bibinfo{author}{\bibfnamefont{W.~D.}
  \bibnamefont{Phillips}}, \bibinfo{journal}{Phys. Rev. A}
  \textbf{\bibinfo{volume}{67}}, \bibinfo{pages}{051603(R)}
  (\bibinfo{year}{2003}).

\bibitem[{\citenamefont{Petsas et~al.}(1994)\citenamefont{Petsas, Coates, and
  Grynberg}}]{Petsas1994a}
\bibinfo{author}{\bibfnamefont{K.}~\bibnamefont{Petsas}},
  \bibinfo{author}{\bibfnamefont{A.}~\bibnamefont{Coates}}, \bibnamefont{and}
  \bibinfo{author}{\bibfnamefont{G.}~\bibnamefont{Grynberg}},
  \bibinfo{journal}{Phys. Rev. A} \textbf{\bibinfo{volume}{50}},
  \bibinfo{pages}{5173} (\bibinfo{year}{1994}).

\bibitem[{\citenamefont{Ashcroft and Mermin}(1976)}]{Mermin1976}
\bibinfo{author}{\bibfnamefont{N.~W.} \bibnamefont{Ashcroft}} \bibnamefont{and}
  \bibinfo{author}{\bibfnamefont{N.~D.} \bibnamefont{Mermin}},
  \emph{\bibinfo{title}{Solid State Physics}} (\bibinfo{publisher}{W.B.
  Saunders Company}, \bibinfo{year}{1976}).

\bibitem[{\citenamefont{Petsas et~al.}(1999)\citenamefont{Petsas, Trich\'e,
  Guidoni, Jurczak, Courtois, and Grynberg}}]{Petsas1999a}
\bibinfo{author}{\bibfnamefont{K.}~\bibnamefont{Petsas}},
  \bibinfo{author}{\bibfnamefont{C.}~\bibnamefont{Trich\'e}},
  \bibinfo{author}{\bibfnamefont{L.}~\bibnamefont{Guidoni}},
  \bibinfo{author}{\bibfnamefont{C.}~\bibnamefont{Jurczak}},
  \bibinfo{author}{\bibfnamefont{J.-Y.} \bibnamefont{Courtois}},
  \bibnamefont{and} \bibinfo{author}{\bibfnamefont{G.}~\bibnamefont{Grynberg}},
  \bibinfo{journal}{Europhys. Lett.} \textbf{\bibinfo{volume}{46}},
  \bibinfo{pages}{18} (\bibinfo{year}{1999}).

\bibitem[{\citenamefont{Grimm et~al.}(2000)\citenamefont{Grimm, Weidemüller,
  and Ovchinnikov}}]{Grimm2000a}
\bibinfo{author}{\bibfnamefont{R.}~\bibnamefont{Grimm}},
  \bibinfo{author}{\bibfnamefont{M.}~\bibnamefont{Weidemüller}},
  \bibnamefont{and}
  \bibinfo{author}{\bibfnamefont{Y.}~\bibnamefont{Ovchinnikov}},
  \bibinfo{journal}{Adv. At. Mol. Opt. Phys.} \textbf{\bibinfo{volume}{42}},
  \bibinfo{pages}{95} (\bibinfo{year}{2000}).

\bibitem[{\citenamefont{Guidoni and Verker}(1999)}]{Guidoni1999a}
\bibinfo{author}{\bibfnamefont{L.}~\bibnamefont{Guidoni}} \bibnamefont{and}
  \bibinfo{author}{\bibfnamefont{P.}~\bibnamefont{Verker}},
  \bibinfo{journal}{J. Opt. B: Quantum Semiclass. Opt.}
  \textbf{\bibinfo{volume}{1}}, \bibinfo{pages}{R23} (\bibinfo{year}{1999}).

\end{thebibliography}
\end{document}